\documentclass[onecolumn,showpacs,preprintnumbers,amsmath,amssymb,PRB]{revtex4}

\usepackage{setspace}
\usepackage{graphicx}
\usepackage{latexsym}
\usepackage{exscale}
\usepackage{amssymb}
\usepackage{amsmath}
\usepackage{dcolumn}
\usepackage{color}
\usepackage{gensymb}

\usepackage{fancyhdr}
\newcommand{\Ket}[1] {\ensuremath{\left| {#1} \right\rangle } }

\allowdisplaybreaks

\begin{document}

\thispagestyle{fancy}

\title[Electron-ion coupling beyond FGR]{On the electron-ion coupling in semiconductors beyond Fermi's Golden Rule}

\author{Nikita Medvedev\footnote{Email: nikita.medvedev@fzu.cz}}
\affiliation{Center for Free-Electron Laser Science, Deutsches Elektronen-Synchrotron DESY, Notkestrasse 85, D-22607 Hamburg, Germany}
\affiliation{Department of Radiation and Chemical Physics, Institute of Physics, Czech Academy of Sciences, Na Slovance 2, 182 21 Prague 8, Czech Republic}
\affiliation{Laser Plasma Department, Institute of Plasma Physics, Czech Academy of Sciences, Za Slovankou 3, 182 00 Prague 8, Czech Republic}

\author{Zheng Li}
\affiliation{SLAC National Accelerator Laboratory, 2575 Sand Hill Road, Menlo Park, CA 94025, USA}
\affiliation{Center for Free-Electron Laser Science, Deutsches Elektronen-Synchrotron DESY, Notkestrasse 85, D-22607 Hamburg, Germany}

\author{Victor Tkachenko} 
\affiliation{Center for Free-Electron Laser Science, Deutsches Elektronen-Synchrotron DESY, Notkestrasse 85, D-22607 Hamburg, Germany}

\author{Beata Ziaja}
\affiliation{Center for Free-Electron Laser Science, Deutsches Elektronen-Synchrotron DESY, Notkestrasse 85, D-22607 Hamburg, Germany}
\affiliation{Institute of Nuclear Physics, Polish Academy of Sciences, Radzikowskiego 152, 
31-342 Krak\'ow, Poland}

\begin{abstract}
In the present work, a theoretical study of electron-phonon (electron-ion) coupling rates in semiconductors driven out of equilibrium is performed. Transient change of optical coefficients reflects the band gap shrinkage in covalently bonded materials, and thus, the heating of atomic lattice. Utilizing this dependence, we test various models of electron-ion coupling. The simulation technique is based on tight-binding molecular dynamics. Our simulations with the dedicated hybrid approach (XTANT) indicate that the widely used Fermi's Golden Rule can break down describing material excitation on femtosecond timescales. In contrast, dynamical coupling proposed in this work yields a reasonably good agreement of simulation results with available experimental data.
\end{abstract}
\pacs{63.20.kd, 64.70.D-, 42.65.Re, 52.50.Jm, 05.70.Ln}
\keywords{Electron-phonon coupling, Fermi's Golden Rule, Silicon, XTANT}
\date{\today}
\maketitle


\section{Introduction}

Since the last decade, various reports on experimental studies of electron-ion coupling in highly-excited materials point to their striking disagreement with theoretical predictions~\cite{Ng2012, Gorbunov2014a}. Different theoretical models overestimate the experimentally observed heating rates by a factor of up to a few orders of magnitude. The rates do then often need to be adjusted to the experimental data~\cite{Lin2008, Gorbunov2014a}. Such a disagreement motivated A. Ng to list the problem of inhibited electron-ion coupling among the 'outstanding questions' in the field of warm dense matter~\cite{Ng2012}. This problem, however, is not limited to the warm-dense-matter regime only, but affects also laser-excited solids which indicates its general significance. Different extensions of the standard electron-phonon coupling approach have been  proposed recently to improve its agreement with experimental data in metals, see, e.g., Refs.~\cite{Petrov2013,Waldecker2016,Migdal2016,Cho2016}. In semiconductors or dielectrics, the details of electron-phonon coupling are even less known, in spite of devoted theoretical efforts such as reported, e.g. , in Refs.~\cite{Fischetti1993, Inogamov2011, Gorbunov2013}.

Experimentally, it is not straightforward to monitor the electron-ion coupling rate with a sufficient temporal resolution since individual electron-ion scattering acts occur at femtosecond timescales~\cite{Celliers1992, White2012}. Most of the experimental studies only infer average heating rates without following their temporal evolution during the relaxation stages of a strongly excited system~\cite{Celliers1992, White2012, Hostetler1999, Hopkins2007, ZhuLi-Dan}.

A system under irradiation with an ultrashort laser pulse undergoes a sequence of excitation and relaxation processes. Electrons after photoabsorption are transiently excited, forming a nonequilibrium distribution which is typically relaxing at femtosecond timescales~\cite{Rethfeld2002}. The thermalized electrons retain high temperature. They then exchange their energies with ions. Significant heating of ions via electron-phonon coupling requires typically a timescale of a few picoseconds~\cite{Lorazo2006}. At the same time, in covalently bonded semiconductors electronic excitation can lead to nonthermal phase transitions (such as melting) induced by the modification of the interatomic potential for sufficiently high fluences~\cite{Stampfli1992, Jeschke1999, Medvedev2013e}. After the local equilibration of electron and ion temperature, ionic system is still hot. Further relaxation may lead to observable material modifications: structural phase transitions, formation of warm dense matter or even plasma, ablation etc.~\cite{Sokolowski-Tinten1995,Fletcher2015,Zhigilei2009,Medvedev2015}.
The stage when the exchange of energy between the thermalized electrons and lattice occurs is of the main interest for the present study.

In this work we test the recently developed tight-binding-based approach for nonadiabatic coupling between electrons and atoms~\cite{Medvedev2015c}. In particular, we analyze the applicability of the Fermi's Golden Rule (FGR) to materials excited at femtosecond timescales, and compare it to a more general dynamical coupling (DC) scheme within the first order approximation~\cite{Landau1976}. It was indicated, for example, in Ref.~\cite{Vorberger2010} that FGR may overestimate electron-ion coupling rates in plasma. Here, we demonstrate that FGR, in general, is inapplicable for the description of electron-ion coupling at femtosecond timescales, i.e., shorter than the characteristic timescales of the electron-phonon interaction in solids.

FGR approach is widely used in the studies of femtosecond electron kinetics in irradiated solids. It was employed in a variety of approaches such as two-temperature model and its extensions~\cite{Lin2008,Gorbunov2014a}; Boltzmann equation~\cite{Allen1987,Rethfeld2002, Pietanza2007,Shcheblanov2012, Mueller2013}; Semiconductor Bloch Equations (SBEs)~\cite{Haas1995, Shah1999, Rossi2002}, and others. Therefore, it is of a crucial importance to be aware of its applicability restrictions. At ultrashort timescales one can use instead the dynamical description appropriate in this regime.

It has been known for some time that transient changes of optical properties may carry information on the undergoing changes of electronic and atomic structure of irradiated materials~\cite{Sokolowski-Tinten1995, Maltezopoulos2008, Gahl2008}. Recently, we pointed out how exactly the pump-probe experiments measuring optical properties of semiconductors can access the information on the electron-ion coupling rates in semiconductors~\cite{Ziaja2015}. This is possible due to the effect of the band gap shrinkage (progressing with the increase of the ion temperature) on the optical properties of the semiconductor. It results in their so-called 'overshooting'. For example, the optical reflectivity may not return to its original value (before the excitation) but raises above it on a picosecond timescale~\cite{Maltezopoulos2008, Gahl2008}. Consequently, with the reflectivity measurements the heating of ions coupled to a hot electronic ensemble can be accessed~\cite{Ziaja2015}. Currently, dedicated  pump-probe experiments can follow transient changes of the optical properties of semiconductors with a few-femtosecond resolution~\cite{Sokolowski-Tinten1995, Harmand2013, Riedel2013}.

\section{Model}

In order to study electron-ion coupling in semiconductors, our hybrid simulation tool XTANT has been used ~\cite{Medvedev2013e, Medvedev2015c}. This recently developed model consists of: (i) a Monte Carlo (MC) code tracing highly excited nonequilibrium electrons emitted after exposure of a material to a free-electron laser pulse; (ii) thermodynamic approach to evaluate low-energy electron distribution within the valence and at the bottom of the conduction band; (iii) Boltzmann collision integral for calculations of electron-ion energy exchange; (iv) tight-binding (TB) molecular dynamics (MD) modeling the atomic motion on a changing potential energy surface. The potential energy surface is affected by the transient state of the electron distribution function, and by the positions of all atoms in the simulation box, with periodic boundary conditions imposed~\cite{Medvedev2013e}.
It also depends on the transient electron eigenstates, and thus changes during the nonthermal phase transition; these effects are implicitly included into the tight binding MD model, as it was discussed in details, e.g., in Ref.~\cite{Medvedev2015}.

In case of optical-pulse  irradiation, photoabsorption proceeds via excitation of valence or conduction-band electrons. Radiation can also be absorbed directly by free carriers. Thus, during the laser  pulse, the electron distribution function looks differently from the discussed case of x-ray irradiation. However, as  it quickly relaxes to the equilibrium  Fermi-Dirac distribution, there is no fundamental difference between the later relaxation of optically or x-ray excited electronic system after the triggering laser pulse and secondary electron cascades are over, provided that the absorbed dose is the same in both cases~\cite{Tkachenko2016}. I.e., also the electron-phonon coupling acts identically in both cases.

Optical properties are calculated from the complex dielectric function (CDF, $\epsilon(\omega)$) within the random-phase approximation following Ref.~\cite{Trani2005}:
\begin{equation}
\epsilon^{\alpha\beta}(\omega)= \delta_{\alpha,\beta} + {{{e}^2\,{\hbar}^2}\over \,{m}^2\Omega\,e_0}\sum_{ij}{{F_{ij}}\over \,E_{ij}^2}\,{f_e(E_{j})-f_e(E_{i})\over \hbar \omega - \hbar \omega_{ij}+i\,{\hbar \gamma}}
\,
\label{rpa}
\end{equation}
where $\Omega$ is the volume of the supercell; $m$ is the mass of a free electron; $\epsilon_0$ is the vacuum permitivity, and $\hbar$ is the Planck's constant.
$\hbar \omega_{ij}$ = $E_{j}$ $-$ $E_{i}$ is transition energy between two eigenstates $\Ket{i}$ and  $\Ket{j}$; 
$f_e(E_{i})$ and $f_e(E_{j})$ are the corresponding transient occupation numbers (electron distribution function); $\omega$ is the frequency as a variable in the complex dielectric function; $F_{ij} = {{|\left\langle i|\hat{p}|j\right\rangle|}^2}$ is the oscillator strength, obtained within Trani's formalism for tight binding calculations~\cite{Trani2005}. As it was demonstrated in Ref.~\cite{Tkachenko2016}, this method can satisfactorily trace transient optical properties of excited solids during phase transitions.

In Eq.(\ref{rpa}), $\gamma$ is the inverse electron relaxation time chosen to be $\gamma = 1.5 \times 10^{13}$ s$^{-1}$~\cite{Tkachenko2016}. Stricktly speaking, in the RPA approximation this parameter should tend to zero, and is kept finite here for numerical purposes only. Its finite value does not affect the results except for  broadening of the peaks in the CDF~\cite{Tkachenko2016}. It is well-known that within the RPA approximation, the electronic collisions cannot be treated consistently~\cite{Mermin1970}. In order to introduce electronic collisions into the model, one has to go beyond the RPA framework (e.g., apply the Lindhard-Mermin dielectric function~\cite{Ropke1999,Arkhipov2014}), which is beyond the scope of the present work.

Nonadiabatic electron-ion coupling is introduced in XTANT via Boltzmann collision integral, $I_{i,j}^{e-at}$:
\begin{widetext}
\begin{eqnarray}{}
\label{Eq:Boltzmann}
I^{e-at}_{i,j} = w_{i,j}
\begin{cases}
f_e(E_i)(2 - f_e(E_j)) - f_e(E_j)(2 - f_e(E_i))G_{at}(E_i - E_j) \ , {\rm for} \ i > j , \\
f_e(E_i)(2 - f_e(E_j))G_{at}(E_j - E_i) - f_e(E_j)(2 - f_e(E_i)) \ , {\rm for} \ i < j ,\
\end{cases}
\label{Fin_coll_int}
\end{eqnarray}
\end{widetext}
where $w_{i,j}$ is the rate for an electron transition between the energy levels $i$ and $j$; here $f_e(E_i)$, a transient electron distribution function, is assumed to be a Fermi-Dirac distribution. It defines electron population on the energy level $E_i$ (eigenstate of the transient TB Hamiltonian); and $G_{at}(E)$ is the integrated Maxwellian function for atoms~\cite{Medvedev2015c}.

Knowledge of the collision integral allows one to evaluate the energy flux between electrons and ions at each time step:
\begin{eqnarray}{}
\label{Eq:Heat_rate}
Q = \sum_{i,j} I_{i,j}^{e-at} \cdot E_{i}
\end{eqnarray}
where the summation is running through all the electronic orbitals for transitions between each pair of levels~\cite{Medvedev2015c}. The transferred energy is then distributed among all the atoms in the simulation box by the appropriate velocity scaling.

\subsection{Electron-ion energy exchange rate}

Let us briefly recall the quantum mechanical theory of the probability of a transition between two states. Generally, the probability of an electron transition between two states (in our case mediated by an ion displacement) during a time $\delta t$ can be written within the first order approximation as follows~\cite{Landau1976}:
\begin{eqnarray}{}
\label{Eq:Landau_P}
P_{i,j} = \left| \left< i (t)| j(t + \delta t) \right> \right|^2,
\end{eqnarray}
where $\left<i (t)| \ \text{and} \ | j(t + \delta t) \right>$ are the $i$-th and $j$-th eigenfunctions of the Hamiltonian at the time instants $t$ and $t+\delta t$, respectively. It reflects the fact that the probability of the transition is determined by the change of the system state during the given time interval. 
To obtain the transition rate, a derivative of this matrix element should be taken, which results in the following expression suitable for finite-difference implementation: 
\begin{eqnarray}{}
\label{Eq:Landau}
w_{i,j} = \left| \left( \langle i(t) | j(t+\delta t) \rangle - \langle i(t +\delta t) | j(t) \rangle \right)/2 \right|^2\frac{1}{\delta t}, 
\end{eqnarray}
Details of the derivation of this expression are given in Appendix A.


If the Hamiltonian is time-independent, the eigenfunctions can be expressed as plane-waves without explicit time-dependence. Under the assumption of  periodic atomic motion within a harmonic potential with the frequency of $\omega_{ph}$ (phonon frequency), the time-derivative of Eq.~(\ref{Eq:Landau_P}) yields the following scattering rate:

\begin{eqnarray}{}
\label{Eq:Dirac}
w_{i,j} = \frac{2}{\hbar^2} |M_{e-at}(E_i,E_j)|^2 \frac{\sin( (\omega_{i j}-\omega_{ph}) \delta t)}{\omega_{i j}-\omega_{ph}},
\end{eqnarray}
where again $\omega_{i j} = (E_i - E_j)/\hbar$ and the matrix element  for electron-ion scattering $M_{e-at}(E_i,E_j) = \left( \langle i(t) | j(t+\delta t) \rangle - \langle j(t) | i(t+\delta t) \rangle \right) (E_j - E_i)/2$. It was derived in Ref.~\cite{Medvedev2015c} and includes the overlap of electronic wave-functions. In case of harmonic ion displacement of small amplitude around the equilibrium positions, $M_{e-at}$ reduces to the conventional form of electron-phonon coupling matrix element in the Debye-H\"uckel form~\cite{Ashcroft1976}.

Note that Eq.~(\ref{Eq:Dirac}) has an explicit time-dependence within the sine function.  This time dependence can be erased, assuming an instant scattering event (Markov process). Physically, this assumption holds at timescales much longer than the duration of an individual scattering event, by setting $\delta t \to \infty$. In this case, the probability reduces to the well-known Fermi's Golden Rule~\cite{Landau1976}:
\begin{eqnarray}{}
\label{Eq:FGR}
w_{i j} = \frac{2 \pi}{\hbar^2} |M_{e-at}(E_i,E_j)|^2 \delta(\omega_{i j}-\omega_{ph}),
\end{eqnarray}
with $\delta(x)$ being the Dirac's delta-function.

However, for the femtosecond timescales we consider here, e.g., in case of femtosecond-laser-pulse irradiation, the assumptions leading to the FGR are not satisfied because~\cite{Medvedev2015}:

(i) In case of strong electronic excitation, atoms might experience rapid modifications of the potential energy surface, leading to an anharmonic atomic motion and significant atom displacements. In extreme case, they can induce nonthermal melting or solid-to-solid phase transition, breaking the crystal symmetry~\cite{Medvedev2015}. That implies that the periodic harmonic motion (phonons, Eq.~(\ref{Eq:Dirac})) cannot be assumed.

(ii) Each individual electron-phonon scattering event in a solid lasts for a time span that can be estimated by the inverse phonon frequency (phonon absorption or emission which takes, typically, a few tens to a hundred of femtoseconds $t_{ph} \sim 1/\omega_{ph}$)~\cite{VanHove1954,March1991,Volkov1998}. Thus, an instant electron-phonon collision (FGR, Eq.~(\ref{Eq:FGR})) cannot be assumed, as was also reported earlier, see e.g.~\cite{Ringhofer2004}.

Therefore, strictly speaking, neither phononic approximation, nor the Fermi's Golden Rule is applicable for modeling of the material excitation induced by intense femtosecond laser  pulses, and the general Eqs.~(\ref{Eq:Landau_P}) or~(\ref{Eq:Landau}) must be used instead. Consequently, the electron transition rates should be calculated at each time step, accounting for the ongoing changes in the atomic subsystem. The corresponding time interval in Eq.~(\ref{Eq:Landau}), $\delta t$ should be set equal to the time step used in the simulation. It is typically on an attosecond time-scale, as will be discussed in the next section.

The dynamical coupling rate, Eq.~(\ref{Eq:Landau}), includes explicit dependence on time without additional assumptions of long timescales and harmonic atomic oscillations. It thus accounts for the evolution of the system during an ongoing individual 'collision' and the induced energy exchange. 

Although FGR is convenient for estimation of transition probabilities (since it does not have an explicit time dependence, and thus does not require time-dependent calculations), it can overestimate the transition rates on short timescales. 
Application of FGR to describe electron-phonon scattering in an electronic system out of equilibrium undergoing rapid changes on a femtosecond timescale, i.e., beyond the limit of validity of FGR, could be one of the reasons for the disagreement of theoretical results with the 'inhibited' electron-ion coupling observed experimentally~\cite{Ng2012,Vorberger2010,White2012}.

Note, however, that Eq.~(\ref{Eq:Landau}) does not have the energy conservation built-in. Electrons do not have to populate only the electronic orbitals (energy levels in Eqs.~(\ref{Eq:Boltzmann},\ref{Eq:Landau})), but can also transiently depart from them. In order to incorporate the full dynamics, the time-dependent Schr\"{o}dinger equation should be used. However, it is computationally too costly for implementation. Thus, the electrons are assumed to always populate the electronic orbitals, and only transitions between them are allowed in the simulation, similar to {\em ab-initio} femtochemical models~\cite{Tully1990,Tully2012}.

For wide-band-gap materials (e.g. diamond in the present study), electron transitions across the band gap are excluded from the collision integral Eq.~(\ref{Eq:Boltzmann}). This is a standard technique used in the {\em ab-initio} femtochemistry~\cite{Zhu2004,Cheng2008}. This restriction is necessary, as an atomic system cannot accept arbitrarily large amount of energy due to electronic transition. This would violate the energy and momentum conservation. In the present calculations, the transitions between the energy levels separated by more than 5 eV are excluded. The simulation results are not sensitive to the particular choice of this acceptance window within the range from $\sim 3$ eV up to $\sim 6$ eV in case of diamond. For the acceptance windows larger than $\sim 6$ eV, some transitions across the band gap start to contribute which makes the coupling in diamond too fast. For the acceptance windows smaller  than $\sim 3$ eV, some transitions within the bands are missing, artificially slowing down the energy exchange. Thus, 5 eV window is used throughout the work for diamond. In case of silicon and graphite, which have small band gaps, the choice of the cut-off for the acceptance window does not affect the predictions, obtained for the excitation and the corresponding electron temperature regime studied here. The same acceptance windows of 5 eV is used for all materials.

\section{Results}

\subsection{Convergence study}

\begin{figure}
  \centering
    \includegraphics[width=0.4\textwidth, trim=10 10 20 20]{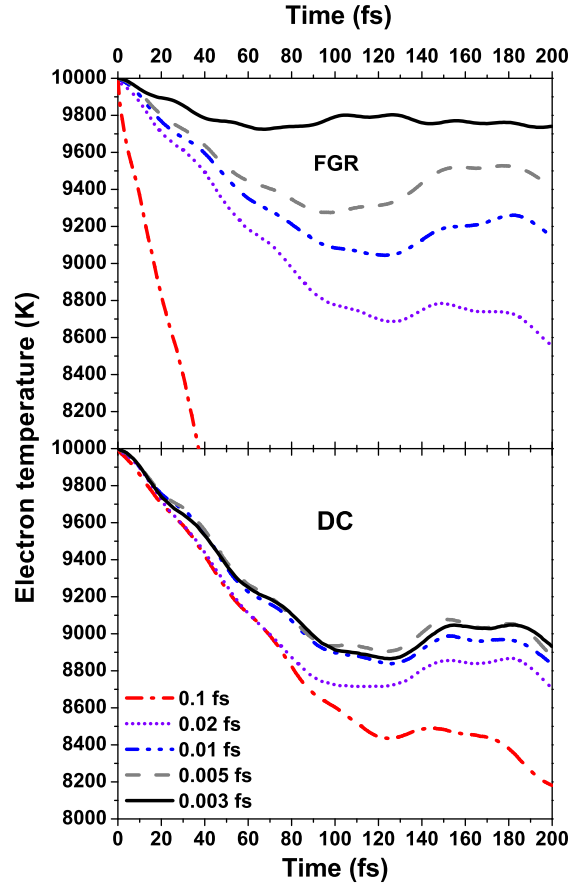}
    \caption[width=0.4\textwidth]{(color online) Convergence study for transient changes of electronic temperature  due to the electron-ion energy exchange in silicon. The changes were calculated with Fermi's Golden Rule (FGR, Eq.~(\ref{Eq:FGR}), top panel) or dynamical coupling approach (DC, Eq.~(\ref{Eq:Landau}), bottom panel). Results obtained with different time-steps used are compared.}
  \label{Pic:200fs}
\end{figure}

In the calculations performed with XTANT model, we noticed that using the FGR, Eq.~(\ref{Eq:FGR}), produces strongly non-convergent results: the heating rate calculated depends on the time-step chosen. This directly contradicts the assumptions necessary for the derivation of the FGR.

Starting from the time-steps of 0.1 fs down to 3 as, no convergence was found in case of FGR calculations, as illustrated in Fig.~\ref{Pic:200fs} for silicon. Here, in all cases the initial electron temperature was set to $T_e = 10000$ K and the ion temperature $T_i = 300$ K. 

The FGR calculations do not converge due to the fact that the assumptions of time-independence of the Hamiltonian and the corresponding wave-functions (points (i) and (ii) in the previous section), necessary for the validity of FGR,  break down. The system's state is changing at each time-step, which makes the matrix element $M_{e-at}$ time-dependent.

In contrast, the calculations using the dynamical coupling (DC) expression for electron-ion coupling, Eq.~(\ref{Eq:Landau}), converged for the time-steps of $\sim 0.01$ fs, see also Fig.~\ref{Pic:200fs}. Discrepancies for the calculated electron temperatures with the time-steps of below $\sim 0.02$ fs are marginal.

The results suggest that our earlier work~\cite{Medvedev2015c} which used FGR with the time-step of 0.1 fs might have underestimated the timescales of the thermal melting of silicon. It seems to take longer than previously reported; however, the values of the damage threshold and the conclusions drawn remain unaffected by the modified timescales (see Appendix B). That is because the damage depends on the total deposited energy, which is the same in both cases. Without additional channels for energy loss, such as heat diffusion, which were not included in the model, a change of the timescale for energy delivery into the atomic system does not affect the predicted damage threshold.

\begin{figure}
  \centering
    \includegraphics[width=0.5\textwidth, trim=10 10 20 20]{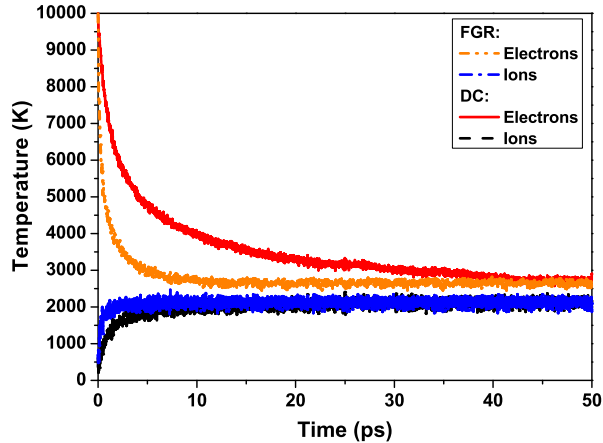}
    \caption[width=0.45\textwidth]{(color online) Evolution of electron and ion temperatures in silicon calculated with electron-ion coupling obtained with Fermi's Golden Rule (FGR, Eq.~(\ref{Eq:FGR}), using time-step of 0.1 fs), and the dynamical coupling (DC) approach, Eq.~(\ref{Eq:Landau}).}
  \label{Pic:FGR_vs_DGR}
\end{figure}

To compare the relaxation timescales, Fig.~\ref{Pic:FGR_vs_DGR} shows the electron-ion thermalization for silicon calculated within the FGR (Eq.~(\ref{Eq:FGR}) using the time-step of 0.1 fs) and the DC (Eq.~(\ref{Eq:Landau}) with the time-step of 0.01 fs for convergence). The FGR calculations with the time-step of 0.1 fs as used in the previous work, Ref.\cite{Medvedev2015c}, are shown here only for comparison.  We can see that the FGR (with the time-step of 0.1 fs) significantly overestimates the timescale for electron-ion energy relaxation. 

In both simulations, the equilibration of temperatures is not exact, as it was also noted in Ref.\cite{Medvedev2015c}. This is due to the finite band gap in silicon, and to the available numerical accuracy. The same effect was observed for diamond and graphite (not shown). 

All further calculations reported below are performed using the dynamical coupling approach with the time-step of 0.01 fs.

\subsection{Effect of electron-ion coupling on optical properties}

In order to compare the modeled dynamical coupling to experimental data, the effect of the electron-ion coupling on the optical properties is analyzed. The calculations are performed for the following pulse and dose parameters in irradiated silicon, corresponding to the pump-probe experiments from Ref.~\cite{Sokolowski-Tinten1995}: absorbed dose of 1.3 times the damage threshold (producing $\sim 0.78$ eV/atom~\cite{Medvedev2015c}); gaussian pulse duration of 60 fs FWHM; probe pulse of 625 nm under $70.5$\degree incidence. 

Although the pump-photon energy modeled here is 30 eV, while in Ref.~\cite{Sokolowski-Tinten1995} an optical pump was used in the experiment, we can compare the results of our simulations with the experimental data, expecting only slight deviations during the pulse, as discussed above~\cite{Tkachenko2016}.
Already by the end of the pulse, one can expect that the simulation and the experimental conditions should approach each other, as demonstrated in Ref.~\cite{Tkachenko2016}, thus, a meaningful comparison can be made. This is due to the similar final electronic states of the system achieved after the excitation with both pulses which fluences were adjusted to ensure the same absorbed dose per atom; after the electrons thermalize, the two cases become identical.

For 30 eV photons, only electrons of $<30$ eV energies are excited. Their  relaxation towards the bottom of the conduction band takes only a few collisions and accomplishes within about a femtosecond~\cite{Medvedev2015a}. Note that in Ref.~\cite{Medvedev2015c}, photon energy of 1000 eV was used, for which the electron cascades are significantly longer (few tens to a hundred fs~\cite{Medvedev2015a}).

\begin{figure}
  \centering
    \includegraphics[width=0.42\textwidth, trim=10 45 10 40]{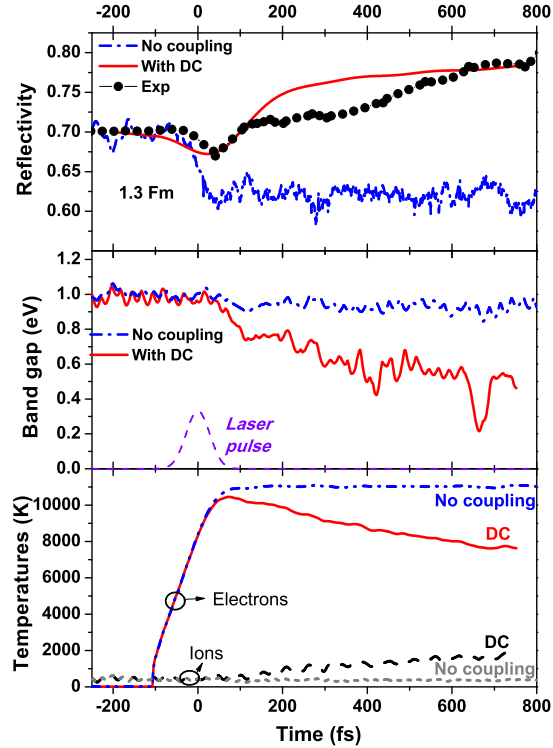}
    \caption[width=0.45\textwidth]{ (color online) {\it Top panel}: optical reflectivity of silicon  at 625 nm probe wavelength under $70.5\degree$ incidence. Calculations with XTANT are shown, using (i) the dynamical coupling approach, Eq.~(\ref{Eq:Dirac}), red line; and (ii) no coupling included (Born-Oppenheimer approximation), blue dash-dotted line. Experimental data from Ref.~\cite{Sokolowski-Tinten1995} are shown for comparison (black circles). {\it Middle panel}: evolution of the band gap of silicon simulated  with and without nonadiabatic electron-ion coupling. Laser  pulse shape is schematically shown with a violet dashed line. {\it Bottom panel}: respective electron and ion temperatures.}
  \label{Pic:Bang_gap}
\end{figure}

Transient values of the optical reflectivity are shown in Fig.~\ref{Pic:Bang_gap}; for better comparison with experimental data, they were smoothed by a convolution with the probe pulse of a finite duration  (60 fs FWHM)~\cite{Sokolowski-Tinten1995, Tkachenko2016}. We can see in Fig.~\ref{Pic:Bang_gap} that the reflectivity calculated with dynamical coupling is reasonably close to the experimental data. 

One can see that the optical reflectivity directly depends on the band gap. When the band gap is shrinking due to the increase of ion temperature (see middle and bottom panels of Fig.~\ref{Pic:Bang_gap}, respectively), the reflectivity is changing accordingly.

Fig.~\ref{Pic:Bang_gap} also demonstrates that using Born-Oppenheimer approximation that excludes electron-ion coupling (marked as 'no coupling' in Fig.~\ref{Pic:Bang_gap}) does not lead to the increase of the ionic temperature, as expected~\cite{Medvedev2015c}, and, correspondingly, does not induce band gap shrinkage (in case of doses below the non-thermal melting threshold). This, in turn, does not trigger any reflectivity increase after the initial drop. The initial drop is caused solely by the electronic excitation.
We do not show here the results with FGR as they did not converge.

These results confirm the idea presented in Ref.~\cite{Ziaja2015}: the 'overshooting effect' observed in experimental data~\cite{Maltezopoulos2008, Gahl2008} is a consequence of the ion heating and the resulting band-gap shrinkage. The optical coefficients such as reflectivity or transmission then encode information on the electron-ion coupling in semiconductors. Thus, the pump-probe experimental schemes can be used to extract the information on electron-ion coupling with femtosecond resolution. Especially, the case of FEL-pump optical-probe scheme would be of interest, allowing for a precise control of the dose absorbed during the irradiation~\cite{Maltezopoulos2008,Harmand2013,Riedel2013}, and a uniform volumetric heating of a sample.

\begin{figure}
  \centering
    \includegraphics[width=0.45\textwidth, trim=10 50 10 40]{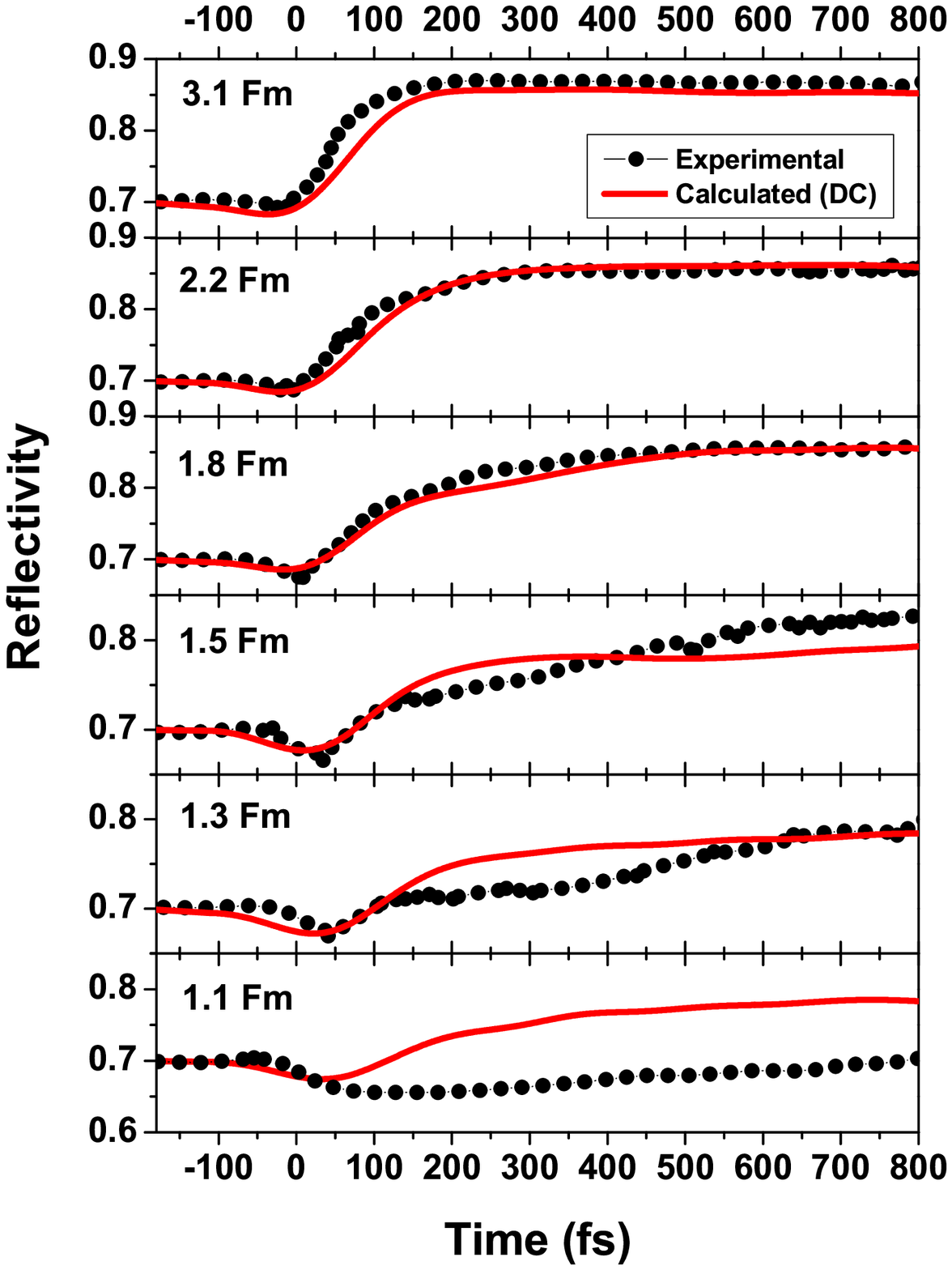}
    \caption[width=0.45\textwidth]{ (color online) Comparison of the calculated and experimental~\cite{Sokolowski-Tinten1995} reflectivity of silicon at 625 nm probe wavelength under $70.5\degree$ incidence for different fluences (in the units of damage threshold, $F_m$). Calculations are performed with the dynamical electron-ion coupling, Eq.~(\ref{Eq:Dirac}). The data were convolved with 60 fs gaussian probe pulse.}
  \label{Pic:Exp_vs_Calc}
\end{figure}

Let us mention here that in case of the absorbed dose above the nonthermal melting threshold, the band-gap collapse can also be induced via nonthermal melting~\cite{Medvedev2013e,Medvedev2015c}. Significant heating of the lattice is then not necessary. For such high doses, the overshooting effect allows to extract time-scales of the predominant nonthermal melting (rather than of the subdominant electron-ion coupling). Such case was studied in in detail in Ref.~\cite{Tkachenko2016}, and will be discussed also here.

The calculated optical reflectivity for different pump-pulse fluences in silicon is shown in Fig.~\ref{Pic:Exp_vs_Calc}. As previously, the data are convolved with a gaussian probe pulse of 60 fs FWHM corresponding to the experimental conditions of Ref.~\cite{Sokolowski-Tinten1995}. Therein,  the fluences are given in the units of the damage threshold, $F_m$. We then compare our results by setting the absorbed dose proportionally to the damage threshold dose, which was previously estimated for Si to be $\sim 0.6$ eV/atom~\cite{Medvedev2015c}. This allows us to avoid any detailed simulations of optical photoabsorption, which is not the focus of the present work.

At high doses above $1.8 F_m$ ($\sim 1.17$ eV/atom) in the present dataset, silicon undergoes nonthermal melting, induced by the modification of the potential energy surface~\cite{Medvedev2015c}. This is an extremely fast process amorphizing the target within the first $\sim 300-500$ fs. It is also affecting the optical reflectivity, as one can see in Fig.~\ref{Pic:Exp_vs_Calc}. Similar behaviour was reported in Ref.~\cite{Tkachenko2016}. The thermal contribution to the transition is here only minor, thus, the electron-ion coupling does not affect much the reflectivity changes. However, a slower melting can be observed for lower doses, because the contribution of thermal effects then becomes more significant~\cite{Medvedev2015c}.

For the doses corresponding to 1.5 and 1.3 $F_m$, the damage proceeds via thermal melting without significant nonthermal effects. It can then last up to a few picoseconds (compare Fig.~\ref{Pic:FGR_vs_DGR}). Our result obtained for the lowest dose, $1.1 F_m$, shows poor agreement with the experimental data. We can identify at least two reasons for that: 

(i) in the XTANT hybrid model, the valence and conduction band electrons are assumed to be in a mutual thermal equilibrium (following a unified Fermi-Dirac distribution). In the experiment~\cite{Sokolowski-Tinten1995} this might not have been the case. The temperature nonequilibrium between valence and conduction band can persist longer for lower pulse fluences~\cite{Rethfeld2002}. As reported, for example, in Ref.~\cite{Rethfeld2002}, for electrons out of equilibrium, electron-ion (electron-phonon) coupling can be significantly slower. Thus, it can be expected that the temperature-equilibrium-based calculations overestimate ion heating, correspondingly underestimating the time for the reflectivity increase in case of the lowest dose.

(ii) Periodic boundary conditions are used in the calculation, which confine all the absorbed energy inside a super-cell. In experiment, this was not the case: heat diffusion may bring some energy out of the irradiated spot. This effect is expected to be more significant at lower  fluences, since in this case the damage process takes longer time.

The two assumptions discussed above have always been  used in our model, for all studied cases of electron-ion coupling: DC, FGR, or in the scenario with no nonadiabatic coupling. Detailed investigation of the effects beyond these assumptions warrants separate dedicated studies, which are out of the scope of the present work.

\section{Discussion}

\begin{figure}
  \centering
    \includegraphics[width=0.5\textwidth, trim=10 20 10 0]{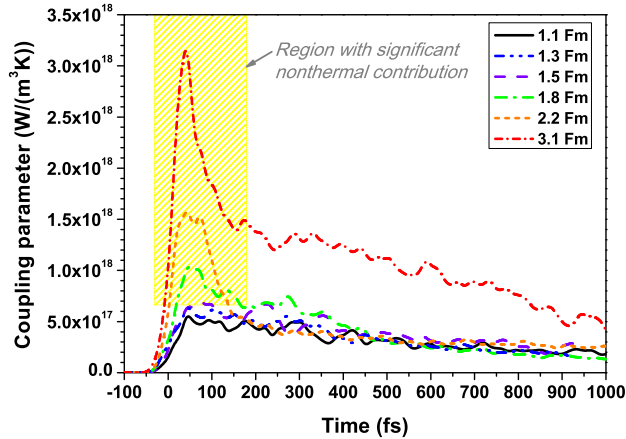}
    \caption[width=0.45\textwidth]{ (color online) Evolution of the dynamical electron-ion coupling parameter (Eq.~(\ref{Eq:Dirac})) in silicon irradiated with different pulse fluences.}
  \label{Pic:Gei}
\end{figure}

Fig.~\ref{Pic:Gei} shows the calculated dynamical electron-ion coupling parameter in silicon at different deposited doses, corresponding to Fig.~\ref{Pic:Exp_vs_Calc}.
The electron-ion coupling parameter (defined as $g_{e-at}(T_e,T_a) = Q/[(T_e-T_a)\Omega]$, where $Q$ is the heat rate defined by Eq.~(\ref{Eq:Heat_rate}), and $\Omega$ is the volume of the simulation supercell) is a time-dependent function of both, electron and ion temperatures.

For the cases below the nonthermal threshold (1.5 $F_m$ and lower in Fig.~\ref{Pic:Gei}), an average value of the electron-phonon coupling parameter appears to be on the order of $3 \cdot 10^{17}$ W/(K m$^3$) during the first picosecond, as it can be seen in Fig.~\ref{Pic:Gei}.
It is sinking with the decrease of the electron temperature and drops below  $10^{17}$ W/(K m$^3$) (at $\sim 10$ ps, not shown), which becomes then closer to the values reported in~\cite{Ng2012}, although for a very different parameters range.

The coupling parameter in Fig.~\ref{Pic:Gei} nonlinearly increases with the increase of the deposited dose. It always decreases during the cooling of the electron system. For the deposited doses above $1.8 F_m$, when nonthermal melting starts to play a role, it exhibits a sudden sharp peak during the first 100 fs after the pulse maximum. 
The appearance of this peak is due to the linear dependence of $M_{e-at}$ on the ion velocity, and reflects a strong 'softening' of the interatomic potential, in which atoms are gaining the kinetic energy not only via nonadiabatic electron-ion coupling but also as a result of the nonthermal melting - due to modification of the interatomic potential through electronic excitation. This means that the electron-ion coupling parameter in this range should not be interpreted as the thermal electron-phonon coupling, but it contains also a contribution from the nonthermal transition. To emphasize this, the region of significant nonthermal contribution is highlighted in Fig.~\ref{Pic:Gei} for the deposited doses above $1.8 F_m$. 
During this time, electrons also reach their highest temperatures. Later, the coupling parameter decreases  due to the decrease of the electronic temperature and to the suppression of nonthermal effects. Finally, it approaches an asymptotic value.

Dedicated analysis of a set of simulation runs with different initial electronic and atomic temperatures showed that the electron-ion coupling parameter in silicon scales approximately as $g_{e-at}(T_e,T_a) \sim T_a T_e^3$, within the limits for: (i) the atomic temperature, $T_a$ from the room temperature up to the melting point $\sim 1687$ K, and (ii) for the electronic temperature, $T_e$, between the room temperature and the limit of a purely nonthermal melting $\sim 17$ kK.

Similar calculations were also performed with XTANT for diamond and graphite. The corresponding tight binding parameters for carbon-based materials are given in Ref.~\cite{Medvedev2013e}. 
The simulation results show that in diamond and graphite, an average electron-ion coupling parameters appear to be on the order of  $\sim5\cdot10^{17}$ W/(K m$^3$) and  $\sim10^{17}$ W/(K m$^3$), respectively. The data reported in Ref.\cite{White2014}, where the coupling parameter in graphite was measured and found to be on the order of $\sim10^{16}$ W/(K m$^3$)),  could not be directly compared with the present calculations, since the high excitation in those experiments lead to electron energies up to a few MeV. Treatment of such relativistic electrons  is beyond reach of our current approach. As electrons of MeV energies require long times for their thermalization and cascading (roughly on the order of $\sim35$ picoseconds by nonrelativistic estimation~\cite{Medvedev2015a}), correspondingly delaying the electron-ion coupling, we expect
 that this might be one of the reasons why the coupling values reported there are lower than those in the present calculations. Another possibility discussed in the literature is the effect of the coupling of collective atomic and electronic modes, that can slow down their energy exchange~\cite{Vorberger2010}. Further studies are required to clarify these issues.
\newline

\section{Conclusions}

In this study we confirmed and utilized the earlier observation~\cite{Ziaja2015} that a change of optical reflectivity is a consequence of the band gap shrinkage in covalently bonded materials, induced by the heating of the atomic lattice (in case of thermal electron-ion coupling). Using this effect, we studied in detail electron-ion coupling in semiconductors. Dedicated simulations were performed for silicon crystals driven out of equilibrium by a femtosecond laser pulse.

Fermi's Golden Rule within the phononic approximation (harmonic approximation for the atomic motion) and the proposed alternative approach, dynamical coupling (DC), were analyzed to model the electron-ion coupling in silicon. The observed non-convergence of the FGR calculations indicates that the widely used phononic approximation and the Fermi's Golden Rule seem to be inapplicable to describe  material excitation at femtosecond timescales. This is in contrast to the dynamical coupling which converged for time-steps $\sim 0.01$ fs, and could describe the optical reflectivity measurements with a reasonably good accuracy.

Further studies are required to test the dynamical coupling approach for different materials. However, as the current analysis already indicates that the assumption of an instant scattering event underlying the Fermi's Golden Rule breaks down at femtosecond timescales, and  time dependence has then to enter explicitly the electron-ion exchange rates, the proposed DC approach seems to be a promising  alternative to the FGR approximation at those ultrashort timescales.

\acknowledgements
The authors thank D.O.~Gericke, S.~Gorbunov, R.~Santra, P.~Terekhin, A.E.~Volkov, and J.~Vorberger for illuminating discussions. 
Partial financial support from the Czech Ministry of Education (Grants LG15013 and LM2015083) is acknowledged by N. Medvedev.
Z. Li is grateful to the Volkswagen Foundation for financial support via the Paul-Ewald postdoctoral fellowship.

\appendix 
\onecolumngrid
\section*{Appendix A: Derivation of the dynamical coupling transition rate}

Electronic transition rate $w_{i,j}(t)$ is defined as a time derivative of the transition probability between the levels \Ket{i} and \Ket{j}, $P_{i,j}(t)$~\cite{Sakurai, Landau1976}:
\begin{eqnarray}{}
\label{Eq:Rate_def}
w_{i,j}(t) = \frac{d P_{i,j}(t)}{d t}
\end{eqnarray}
Introducing the transition rate:
\begin{eqnarray}{}
\label{Eq:Rate_def_lim}
w_{i,j}(t) = \lim_{t \to t_0} \frac{(P_{i,j}(t)  - P_{i,j}(t_0))}{t-t_0}
\end{eqnarray}
into the numerical model, we use the finite-difference approximation for the derivative:
\begin{eqnarray}{}
\label{Eq:Rate_def_finite}
w_{i,j}(t) \approx  \frac{(P_{i,j}(t) - P_{i,j}(t_0))}{\delta t}
\end{eqnarray}
where the time step $\delta t \equiv  t - t_0$ has  to be sufficiently short. It corresponds to  the time-step within our TBMD scheme of XTANT.

As we discretize time into sufficiently short time-steps $\delta t$, the interatomic forces are assumed constant during each time-step within TBMD part of XTANT. The Hamiltonian does not change during the time-step, being updated only at the beginning of the next step. This implies that any time-dependent perturbing potential can be represented  as a sequence of step-like functions: constant within any time-step and updated before the next one. The corresponding matrix element then reads:
\begin{eqnarray}{}
\label{Eq:V_def_const}
V_{i,j}(t) = V_{i,j}\theta(t-t_0)
\end{eqnarray}
where $\theta(x)$ is Heaviside step-function, and $V_{i,j} \equiv V_{i,j}(t_0)$ is constant during $\delta t$. The  perturbation is switched on instantaneously at time $t_0$.  Consequently,  $P_{i,j}(t_0) = 0$. The transition rate then reads:
\begin{eqnarray}{}
\label{Eq:Rate_def_finite}
w_{i,j}(t) = \frac{P_{i,j}(t)}{\delta t}
\end{eqnarray}
We apply the approximate transition probability (for instanteneous approximation consistent with our scheme here) derived in Ref.~\cite{Landau1976} (Eq.~(41.5) therein):
\begin{eqnarray}{}
\label{Eq:P_solved}
P_{i,j}(t) = \left| \frac{ V_{i,j} }{h \omega_{i,j}} \right|^2,
\end{eqnarray}
where $\omega_{i,j} = (E_i - E_j)/\hbar$, and $E_i$ are the transient electronic energy levels (band structure). With  Eq.~(\ref{Eq:Rate_def_finite}) we arrive at the transition rate:
\begin{eqnarray}{}
\label{Eq:w_solved}
w_{i,j}(t) = \frac{1}{\delta t} \left| \frac{ V_{i,j} }{h \omega_{i,j}} \right|^2
\end{eqnarray}
Utilizing our previously derived finite-difference expression for the perturbation matrix element from Ref.~\cite{Medvedev2015c}: $V_{i,j} = \left( \langle i(t_0) | j(t_0+\delta t) \rangle - \langle i(_0t+\delta t) | j(t_0) \rangle \right) (E_j - E_i)/2$, we obtain the final expression for the transition rate that has been implemented into XTANT code:
\begin{eqnarray}{}
\label{Eq:w_fin}
w_{i,j}(t) = \frac{1}{\delta t} \left| \left[ \langle i(t_0) | j(t_0+\delta t) \rangle - \langle i(t_0+\delta t) | j(t_0) \rangle \right]/2 \right|^2
\end{eqnarray}

\section*{Appendix B: calculated damage thresholds}

Using the dynamical coupling rate from Eq.~(\ref{Eq:Landau}) and the FGR Eq.~(\ref{Eq:FGR}) (with the time-step of 0.1 fs), the following results were obtained for damage thresholds. We sum them up in Tables~\ref{Tab:Diamond} and~\ref{Tab:Silicon}.

\begin{table}[th]
\centering
\caption{Damage thresholds and transition timescales in diamond calculated with: (i) Born-Oppenheimer (BO) approximation that excludes nonadiabatic coupling; (ii) Fermi's Golden Rule (FGR) Eq.~(\ref{Eq:FGR}) with the time-step of 0.1 fs; and (iii) dynamical electron-ion coupling (DC) Eq.~(\ref{Eq:Landau}). Experimental references are also given for comparison wherever available.} 
\label{Tab:Diamond}
 \begin{tabular}{|c|c|c|c|c|}
 \cline{1-5}
  & BO & FGR & DC & Experiment \\ \cline{1-5}
 Damage Threshold, eV/atom & 0.7-0.75~\cite{Medvedev2013,Medvedev2013e} & 1-1.1 & 0.65-0.7 & $\sim 0.7$~\cite{Gaudin2013} \\ \cline{1-5}
 Timescales, fs & $\sim 80-130$ & $\sim 80$ & $\sim 80-150$ & -- \\ \cline{1-5}
 \end{tabular}
\end{table}

\begin{table}[th]
\centering
\caption{Damage thresholds and transition timescales in silicon calculated with: (i) Born-Oppenheimer (BO) approximation that excludes nonadiabatic coupling; (ii) Fermi's Golden Rule (FGR) Eq.~(\ref{Eq:FGR}) with the time-step of 0.1 fs; and (iii) dynamical electron-ion coupling (DC) Eq.~(\ref{Eq:Landau}). Both damage threshold, for thermal and nonthermal melting, are presented. Experimental references are also given for comparison wherever available.} 
\label{Tab:Silicon}
 \begin{tabular}{|c|c|c|c|c|}
 \cline{1-5}
  & BO & FGR & DC & Experiment \\ \cline{1-5}
 Thermal melting (TM), eV/atom & -- & 0.65~\cite{Medvedev2015c} & 0.65 & --  \\ \cline{1-5}
 Timescales for TM, ps & -- & $\sim 1$ & $\sim 2$ & $1.5-2$~\cite{Beye2010,Harb2006} \\ \cline{1-5}
 Nonthermal melting (NTM), eV/atom & 2.1~\cite{Medvedev2015c} & 0.9~\cite{Medvedev2015c} & 0.9 & $\sim 0.9-1$~\cite{Harb2008} \\ \cline{1-5}
 Timescales for NTM, fs & $\sim 300$~\cite{Medvedev2015c} & $\sim 300-500$~\cite{Medvedev2015c} & $\sim 300-1000$ & 300-500~\cite{Sokolowski-Tinten1995,Sundaram2002,Harb2008} \\ \cline{1-5}
 \end{tabular}
\end{table}

In case of dynamical coupling, the damage thresholds are lowered for both silicon and diamond, when compared to the results with  the Born-Oppenheimer approximation. The effect is more noticeable for silicon, while for diamond it is only minor. Transition timescales  with both approaches are similar for diamond, and differing for silicon.

In contrast, FGR coupling in diamond induces much faster heating rates (for the simulation time-step of 0.1 fs), which prevents graphite from forming; this disagrees with the experimental observations of graphitization after an FEL irradiation~\cite{Gaudin2013} at the respective threshold fluence.

For silicon the transition timescales obtained with FGR (again, with the time-step of 0.1 fs to represent earlier works) correspond to those obtained with the Born-Oppenheimer approximation (for the non-thermal melting threshold) and are much shorter than those obtained with dynamical approach. The experimental damage threshold for nonthermal melting in silicon given in Table~\ref{Tab:Silicon} is estimated from the reported threshold density of electron $\sim 6 \%$ in Ref.~\cite{Harb2008}.


\bibliographystyle{ieeetr}
\bibliography{My_Collection}

\end{document}